\documentclass[twocolumn,showpacs,preprintnumbers,superscriptaddress,amsmath,amssymb]{revtex4-1}
\usepackage{graphicx,bm,natbib,placeins,times,enumitem}
\usepackage[rightcaption]{sidecap}
\usepackage[colorlinks=true]{hyperref}

\begin{document} 

\title{Slim Fractals: The Geometry of Doubly Transient Chaos}

\author{Xiaowen Chen}
\altaffiliation[Currently at: ]
{Department of Physics, Princeton University, Washington Road, Princeton, NJ 08544, USA}
\affiliation{Department of Physics and Astronomy, Northwestern University, Evanston, IL 60208, USA}

\author{Takashi Nishikawa}
\affiliation{Department of Physics and Astronomy, Northwestern University, Evanston, IL 60208, USA}
\affiliation{Northwestern Institute on Complex Systems, Northwestern University, Evanston, IL 60208, USA}

\author{Adilson E. Motter} 
\affiliation{Department of Physics and Astronomy, Northwestern University, Evanston, IL 60208, USA}
\affiliation{Northwestern Institute on Complex Systems, Northwestern University, Evanston, IL 60208, USA}

\begin{abstract}
Traditional studies of chaos in conservative and driven dissipative systems have established a correspondence between sensitive dependence on initial conditions and fractal basin boundaries, but much less is known about the relation between geometry and dynamics in undriven dissipative systems. These systems can exhibit a prevalent form of complex dynamics, dubbed doubly transient chaos because not only typical trajectories but also the (otherwise invariant) chaotic saddles are transient. This property, along with a manifest lack of scale invariance, has hindered the study of the geometric properties of basin boundaries in these systems---most remarkably, the very question of whether they are fractal across all scales has yet to be answered. 
Here we derive a general dynamical condition that answers this question, which we use to demonstrate that the basin boundaries can indeed form a true fractal; in fact, they do so generically in a broad class of transiently chaotic undriven dissipative systems.
Using physical examples, we demonstrate that the boundaries typically form a {\it slim fractal}, which we define as a set whose dimension at a given resolution decreases when the resolution is increased.
To properly characterize such sets, we introduce the notion of \textit{equivalent dimension} for quantifying their relation with sensitive dependence on initial conditions at all scales. 
We show that slim fractal boundaries can exhibit complex geometry even when they do not form a true fractal and fractal scaling is observed only above a certain length scale at each boundary point.
Thus, our results reveal slim fractals as a geometrical hallmark of transient chaos in undriven dissipative systems.
\end{abstract}

\onecolumngrid\hfill 
\href{https://doi.org/10.1103/PhysRevX.7.021040}{\small Phys. Rev. X {\bf 7}, 021040 (2017)}
\bigskip\twocolumngrid

\maketitle

\section{Introduction}

Physicists often relate chaos with fractal basin boundaries and sensitive dependence on initial conditions~\cite{ott,telbook,Aguirre:2009,leaking2013,review2015}. While the former is a geometrical concept and the latter is inherently dynamical, the correspondence between the two has been established for conservative systems and driven dissipative systems. 
For example, in driven dissipative systems, the geometry and dynamics of a chaotic attractor are explicitly related through the Kaplan-Yorke formula~\cite{Kaplan:1979}, which connects the information dimension of the attractor with its Lyapunov exponents.
A generalization of this formula to chaotic saddles is the Kantz-Grassberger relation~\cite{KG}, which connects the information dimensions along unstable directions with the associated Lyapunov exponents and the overall rate of escape from the saddle.
While some fundamental open problems remain subjects of active research (e.g., the properties and applications of transient chaos~\cite{Hof2008,Altmann2010,Ravasz2011,Moura2011,Wolfrum2011}, as well as the robustness~\cite{Joglekar2014}, the classification~\cite{Sander2015}, and the very definition~\cite{Hunt2015} of chaos), studies of chaos in such systems are relatively mature~\cite{Motter2013}.

In contrast, much less is understood about the relation between dynamics and geometry in the large class of physical processes categorized as dissipative but undriven, in which energy dissipated is not balanced by energy injected into the system.
Examples of such systems abound, including coalescing binary systems in astrophysics, interacting vortices in viscous flows, chemical reactions approaching equilibrium, and many forms of self-organization. 
It also includes various arcade games 
(e.g., pinball) and games of chance (e.g., coin flipping and dice throwing)
as well as cue and throwing sports (e.g., billiards and bowling).
Due to the monotonic decrease of energy to its minima in such systems, all trajectories in a compact phase space will eventually settle to one of the fixed points, and the fixed points are the only invariant sets.
Yet, for a transient period of time the dynamics can be very complicated and demonstrate sensitive dependence on initial conditions.

A recent paper by a collaboration involving one of us \cite{motter13} studied the nature of the {\it dynamics} of such systems. 
It was demonstrated that these systems show fundamentally different properties when compared to driven dissipative systems. 
In particular, they exhibit doubly transient chaos: system trajectories transiently follow a chaotic saddle which is itself transient.
Moreover, the fraction of unsettled trajectories follows a doubly exponential function of time, which corresponds to an exponential settling rate rather than the constant settling rate observed in driven dissipative systems.
However, the geometry of the attraction basins has not been characterized, and has been generally perceived as a very hard problem to address because these systems do not enjoy scale invariance (i.e., the basin boundaries do not exhibit any form of self-similarity, not even statistically). 
While it is known~\cite{telbook,motter13,review2015} that the attraction basins are intertwined and appear fractal-like, the absence of invariant chaotic saddles suggests that the basin boundaries may be simple at sufficiently small scales.
Hence, the question remains whether the boundaries are true fractals.
If the boundaries are fractals, what leads to the fractality despite the lack of invariant chaotic saddles?
If they are not fractals, is there a characteristic length scale for the system that defines the resolution at which the boundaries become simple? 
How can we quantify the sensitive dependence on initial conditions in terms of their geometry?
What roles do the observation length scale and computational precision play in one's ability to measure and simulate the dynamics of the system?

In this article, we investigate the geometry of attraction basins to address the questions posed above. 
We derive the condition under which the boundaries form a true fractal set (i.e., successive magnifications of the boundaries reveal new structures at arbitrarily small scales) and have the Wada property~\cite{wada} (i.e., any boundary point between two basins is also a boundary point between all basins) for a general class of undriven dissipative systems.
We show that this condition is satisfied generically, indicating that true fractal basin boundaries and the associated sensitive dependence on initial conditions are not only possible but are in fact common.
The boundaries can also form {\it a finite-scale fractal}, characterized at each point by 
a finite length scale above which the fractal property is observed and below which the boundaries are simple around that point.
Through extensive, high-precision numerical simulations on physical examples---the dynamics of a roulette of different shapes---we show that 
this {\it fractality length scale} can be smaller than the resolution typically used in simulations, making such basin boundaries practically indistinguishable from true fractals.  
We also find that, as a function of phase-space position, the fractality length scale can vary across many orders of magnitude.
A common feature shared by the observed fractal and finite-scale fractal basin boundaries is that (at a given phase-space position) the fractal dimension for a given length scale decreases with the decrease of that length scale.
Since this property implies that the boundaries would appear to cover less space when observed at higher resolution, we call such sets {\it slim fractals}.
For characterizing the complex geometry of such boundaries, the existing fractal dimensions are not adequate, whether they are defined asymptotically at zero length scale or defined at a given finite length scale.
Thus, to capture the cumulative effect of fractal scaling across all scales, we define the notion of {\it equivalent dimension} based on the process of increasing the initial-state accuracy to reduce the final-state uncertainty.

In the following, we first introduce the class of systems we consider and derive the condition for the fractality of their basin boundaries (Sec.~II).
We then apply the condition to the roulette systems and numerically validate the results (Sec.~III). 
This is followed by the introduction of the equivalent dimension and its application to the roulette systems (Sec.~IV).
We provide concluding remarks in the final section (Sec.~V).

\section{Fractality condition for basin boundaries}

For concreteness, here we focus on the class of two-dimensional potential systems with frictional dissipation having $n$ stable equilibria symmetrically located around an unstable equilibrium and separated by ``hills'' in the potential function.  The equations of motion for such a system are  
\begin{equation}\label{eqn:main}
\ddot{x} + \mu\dot{x} = -\frac{\partial U}{\partial x},\quad
\ddot{y} + \mu\dot{y} = -\frac{\partial U}{\partial y},
\end{equation} 
where $\mu$ is the dissipation constant and $U(x,y)$ is the potential function.  The dynamics of this system can be regarded as a scattering process, in which a trajectory entering the neighborhood of the unstable equilibrium swings back and forth chaotically between the hills before approaching one of the stable equilibria.  The dynamics is thus dominated by the shape of the potential in this scattering region near the unstable equilibrium, which we define to be the origin.  Writing in polar coordinates, the shape of the potential function near the origin is determined by the leading term in the expansion
\begin{equation}\label{eqn:potential}
U(r,\theta) = a_2(\theta)r^2 + a_3(\theta)r^3 + \cdots,
\end{equation}
if $U(r,\theta)$ is smooth with respect to $r$.
The symmetry of the system implies that the coefficients are $n$-fold periodic functions: $a_i(\theta+\frac{2\pi j}{n})=a_i(\theta)$ for each integer $j$.  
The coefficient $a_2(\theta)$ additionally satisfies $a_2(\frac{2\pi j}{n}) \le 0$ and $a'_2(\frac{2\pi j}{n}) = 0$ for each $j$, because the attracting equilibria can be assumed to be located along the lines $\theta=\frac{2\pi j}{n}$ without loss of generality.  

We establish that {\it the fractality of the basin boundaries is determined by system trajectories that move down a hill in the potential and approach the neighborhood of the origin}.
Specifically, we show that the basin boundaries are: (1) fractal if all such trajectories pass through the neighborhood, and (2) not fractal if some of them can asymptotically approach the origin without passing through it.
Case (1) includes the generic situations in which $a_2(\theta)r^2$ is the leading term in Eq.~\eqref{eqn:potential}, coefficient $a_2(\theta)$ takes both positive and negative values depending on $\theta$ [with positive $a_2(\theta)$ in the direction of the hills], and the dissipation is sufficiently weak.
Case (2) includes the non-generic situation in which $a_2(\theta)$ is identically zero [thus making the leading term in Eq.~\eqref{eqn:potential} cubic or higher] and the leading coefficient $a_j(\theta)$ takes both positive and negative values.

\begin{SCfigure*}[0.55][tb]
\includegraphics[width=0.6\textwidth]{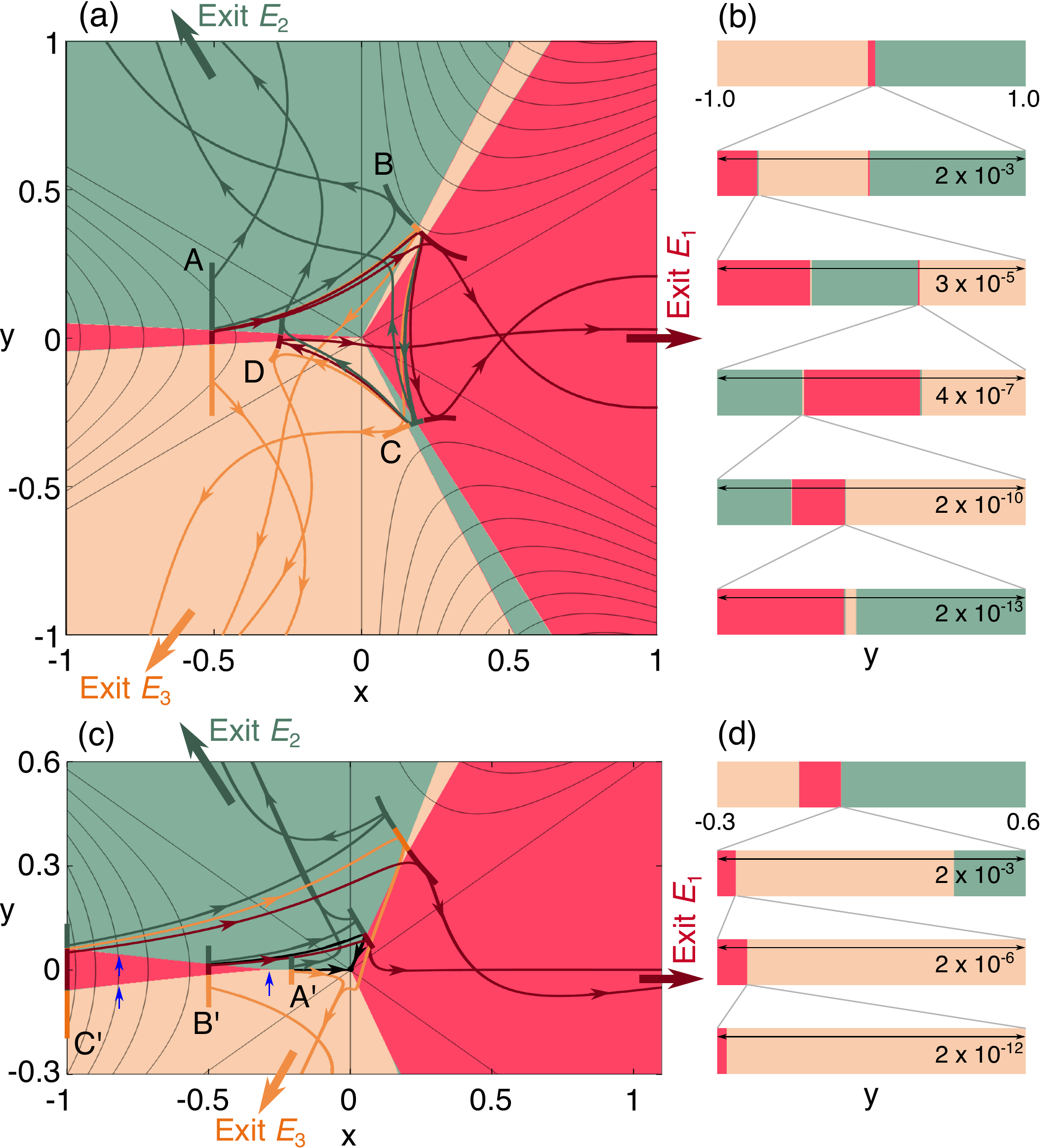}
\caption{Geometry and dynamics in the scattering region of undriven dissipative systems. (a) Fractal basin boundaries in the two-dimensional cross section $\dot{x} = \dot{y} = 0$ for potential $U_2(r,\theta)$ and $\mu=0.2$.  The basins of the exits $E_1$, $E_2$, and $E_3$ are colored red, green, and beige, respectively.  Trajectories starting on the one-dimensional cross section at $x=-0.5$ (vertical line segment A) are shown with arrows indicating the direction of flow and colors indicating the basin to which they belong.
(b) Successive magnifications of the one-dimensional cross section at $x=-0.5$ in (a), showing the fractal nature of the basin boundaries.
(c) Finite-scale fractal basin boundaries for the potential $U_3(r,\theta)$ and $\mu=1$.  Trajectories with initial conditions on the cross sections at $x=-0.2$ (segment A$'$), $x=-0.5$ (segment B$'$), and $x=-1$ (segment C$'$) are shown.  
(d) Successive magnifications of the cross section at $x=-1$ in (c), revealing a simple boundary.\vspace{20mm}}\label{fig:potential_syst}
\end{SCfigure*}

As an example for case (1), consider the potential 
\begin{equation}
U_2(r,\theta) \equiv -r^2\cos 3\theta
\end{equation}
[i.e., $n=3$, $a_2(\theta)=-\cos 3\theta$, and no higher-order terms].
Although this potential makes Eq.~\eqref{eqn:main} an open scattering system with no attractors, it can be regarded as an approximation of a system that has $n=3$ attractors far away from the scattering region.
There are three possible ways [denoted $E_1$, $E_2$, and $E_3$; see Fig.~\ref{fig:potential_syst}(a)] for a trajectory to exit the scattering region.
We can show that, between any two trajectories starting on the vertical line segment labeled A in Fig.~\ref{fig:potential_syst}(a) with velocity zero and eventually leaving the region through two different exits, we can find another trajectory that goes to the third exit (see  Appendix~\ref{sec:full-analysis}). 
Such a situation is illustrated in Fig.~\ref{fig:potential_syst}(a) by the red and green trajectories starting near the color boundary on segment A, which turn around near curve B and exit the region through $E_1$ and $E_2$, respectively. 
We indeed see that the orange trajectory starting between them turns around near B, passes the neighborhood of the origin, and exits through $E_3$.
Since the same situation can occur after an arbitrary number of oscillations between the hills (e.g., after bouncing off B once and reaching C), this translates to the following property of the basins on A: between any two segments of different colors, we can always find a segment of the third color. 
These geometrical properties are verified numerically by successive magnifications near a boundary point in Fig.~\ref{fig:potential_syst}(b).
We note that our argument for segment A (on which the initial velocity is zero) can be extended to an arbitrary line segment in the full four-dimensional phase space connecting points from different basins (see  Appendix~\ref{sec:full-analysis}).
This implies that any cross section of the neighborhood of any boundary point has a similar Cantor-set structure and has the Wada property, establishing that the entire set of basin boundaries is fractal.

\begin{figure}[t]
\includegraphics[width=0.9\columnwidth]{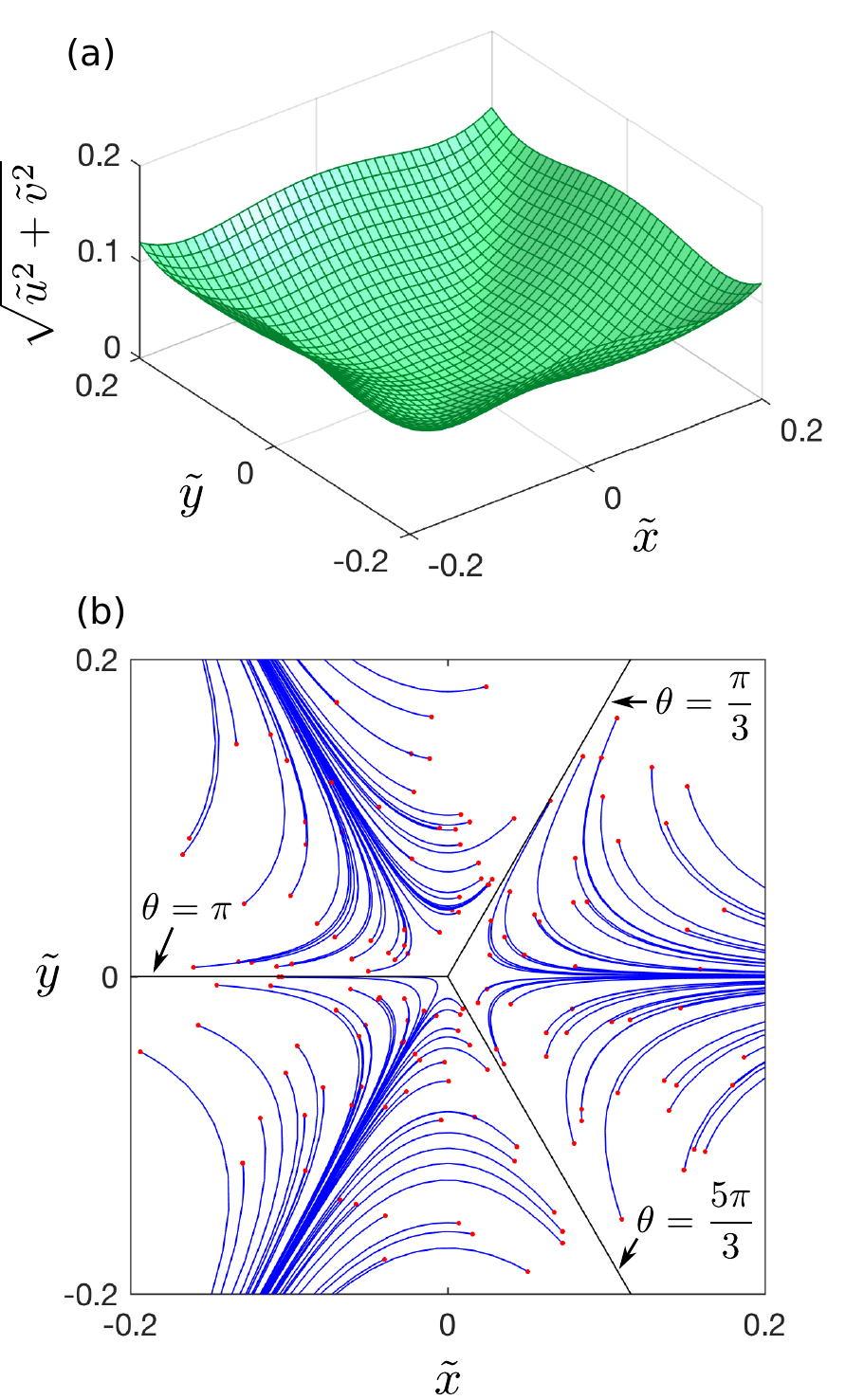}
\caption{Center manifold reduction of the dynamics at the origin for Eq.~\eqref{eqn:main} with $U_3(r,\theta)$.
(a) Three-dimensional projection of the approximate center manifold given by Eq.~\eqref{eq:center_manifold}.
(b) Approximate dynamics on the center manifold, given by Eq.~\eqref{eq:center_manifold_dyn}.
The blue curves are trajectories initiated at the (randomly chosen) initial conditions indicated by red dots.}
\label{u_3_center_manifold_fig}
\end{figure}

As an example for case (2), consider the potential 
\begin{equation}
U_3(r,\theta) \equiv -r^3\cos 3\theta
\end{equation}
[i.e., $n=3$, $a_2(\theta)=0$, $a_3(\theta)=-\cos 3\theta$, and no higher-order terms].
With this potential, Eq.~\eqref{eqn:main} is also an open scattering system that approximates one with three attractors.
In this case, we can show that there exists a finite-length line segment $\{-r_s \le x \le 0,\,\, y=0\}$ from which all trajectories approach the origin asymptotically (see Appendix~\ref{sec:full-analysis}) and that this segment is a simple boundary between the basins of $E_2$ and $E_3$, which does not belong to the boundary of the basin of $E_1$.
This is because any trajectory starting above (below) this segment with zero initial velocity, no matter how close it is to the segment, moves toward the origin initially but soon curves away and exits through $E_2$ ($E_3$).
The trajectories starting exactly on the segment do not exit the region at all.
The green, orange, and black trajectories starting from A$'$ in Fig.~\ref{fig:potential_syst}(c) illustrate this situation.
Thus, every point on this segment is a boundary point between basins of $E_2$ and $E_3$, and hence is a non-Wada point, implying that successive magnifications around this segment would not reveal any finer structures.
We can further show that the segment splits into two branches forming simple boundaries, each of which in turn splits into two branches forming simple boundaries [see branching points indicated by blue arrows in Fig.~\ref{fig:potential_syst}(c)], and so on, composing a binary tree of simple boundary segments.
Thus, the boundaries are not fractal [as numerically verified by successive magnifications in Fig.~\ref{fig:potential_syst}(d)]; however, since they have a Cantor set structure down to finite length scales (which are different for different branches), we say that such boundaries form a {\it finite-scale fractal}. 
We now generalize this result to lift the zero initial velocity assumption.
Our argument is based on applying the center manifold reduction~\cite{Guckenheimer:2013} to the equilibrium at the origin.
Transforming Eq.~\eqref{eqn:main} with $U_3(r,\theta)$ into a suitable coordinate system $(\tilde{x},\tilde{y},\tilde{u},\tilde{v})$, we determine the local center manifold to be 
\begin{equation}\label{eq:center_manifold}
\begin{pmatrix}
\tilde{u}\\
\tilde{v}
\end{pmatrix} = \frac{3}{\mu^2}\begin{pmatrix}
-\tilde{x}^2 + \tilde{y}^2\\
2\tilde{x}\tilde{y}
\end{pmatrix}
\end{equation}
and the dynamics on that manifold to be
\begin{equation}\label{eq:center_manifold_dyn}
\begin{split}
\dot{\tilde{x}} &= \frac{3}{\mu}(\tilde{x}^2 - \tilde{y}^2),\\
\dot{\tilde{y}} &= -\frac{6}{\mu}\tilde{x}\tilde{y},
\end{split}
\end{equation}
up to second order in $\tilde{x}$ and $\tilde{y}$.
These are both visualized in Fig.~\ref{u_3_center_manifold_fig}.
By extending the local basin boundaries determined by Eq.~\eqref{eq:center_manifold_dyn} to the global phase space, we establish that the full set of basin boundaries is a finite-scale fractal (see Appendix~\ref{sec:full-analysis} for details).

It is interesting to note that the stable manifold of just one equilibrium (the origin) is responsible for the full complexity of the basin boundaries---whether they are fractal or finite-scale fractal---for the class of systems we consider.
To see this, note that the basin boundaries consist of all points from which the trajectories never leave the scattering region. 
Since the only possible asymptotic state in this region is the unstable equilibrium at the origin, any trajectory starting from a boundary point must approach the equilibrium.
Conversely, any point from which the trajectory converges to the equilibrium is a boundary point.
This is because one can always find an arbitrarily small change to the initial point that would make the trajectory steer left or right just before converging to the equilibrium, and eventually leave the scattering region through one exit or another.
Thus, the set of boundary points is the stable manifold of the equilibrium.

In addition to case (2) discussed above, finite-scale fractals can arise when the origin is a local maximum of the potential [e.g., when $a_2(\theta)<0$ for all $\theta$], if the higher-order terms in Eq.~\eqref{eqn:potential} create unstable saddle points that play a role similar to that played by the origin in our argument above.  We will see an example of this situation below.
Also, the transition between fractal and finite-scale fractal boundaries can be studied using the class of potentials 
\begin{equation}
U_\alpha(r,\theta) \equiv -r^\alpha\cos 3\theta
\end{equation}
with arbitrary real parameter $\alpha$.
Indeed, we can fully characterize this {\it fractality transition}: the boundaries are fractal if $\alpha \le 2$ and finite-scale fractal if $\alpha > 2$ [see Appendix~\ref{sec:U-alpha} for the analysis and 
Appendix~\ref{sec:frac_trans_U-alpha}
for numerical verification].

Finally, we note that our arguments above do not rely on the linearity of the dissipative term in Eq.~\eqref{eqn:main} and can also be applied to  systems with nonlinear dissipation (i.e., when $\mu$ is not constant and instead depends on the position, such as in  electric circuits with nonlinear resistors~\cite{Semenov:2016} and in nanomechanical resonators~\cite{Imboden:2013}).
In particular, our fractality condition based on the behavior of the trajectories approaching the origin remains valid for any nonnegative function $\mu=\mu(r,\theta)$, and the condition can be expressed in terms of $\mu(r,\theta)$ 
(see Appendix~\ref{sec:U-alpha}).
For instance, if the dissipation is of the form $\mu=\mu_0 r^q$, this condition reads as follows: the boundaries form a true fractal if $\alpha \le 2(1+q)$ and a finite-scale fractal if $\alpha > 2(1+q)$.

\section{Roulette as a model system}

\begin{figure*}[t]
\begin{center}
\includegraphics[width=5.75in]{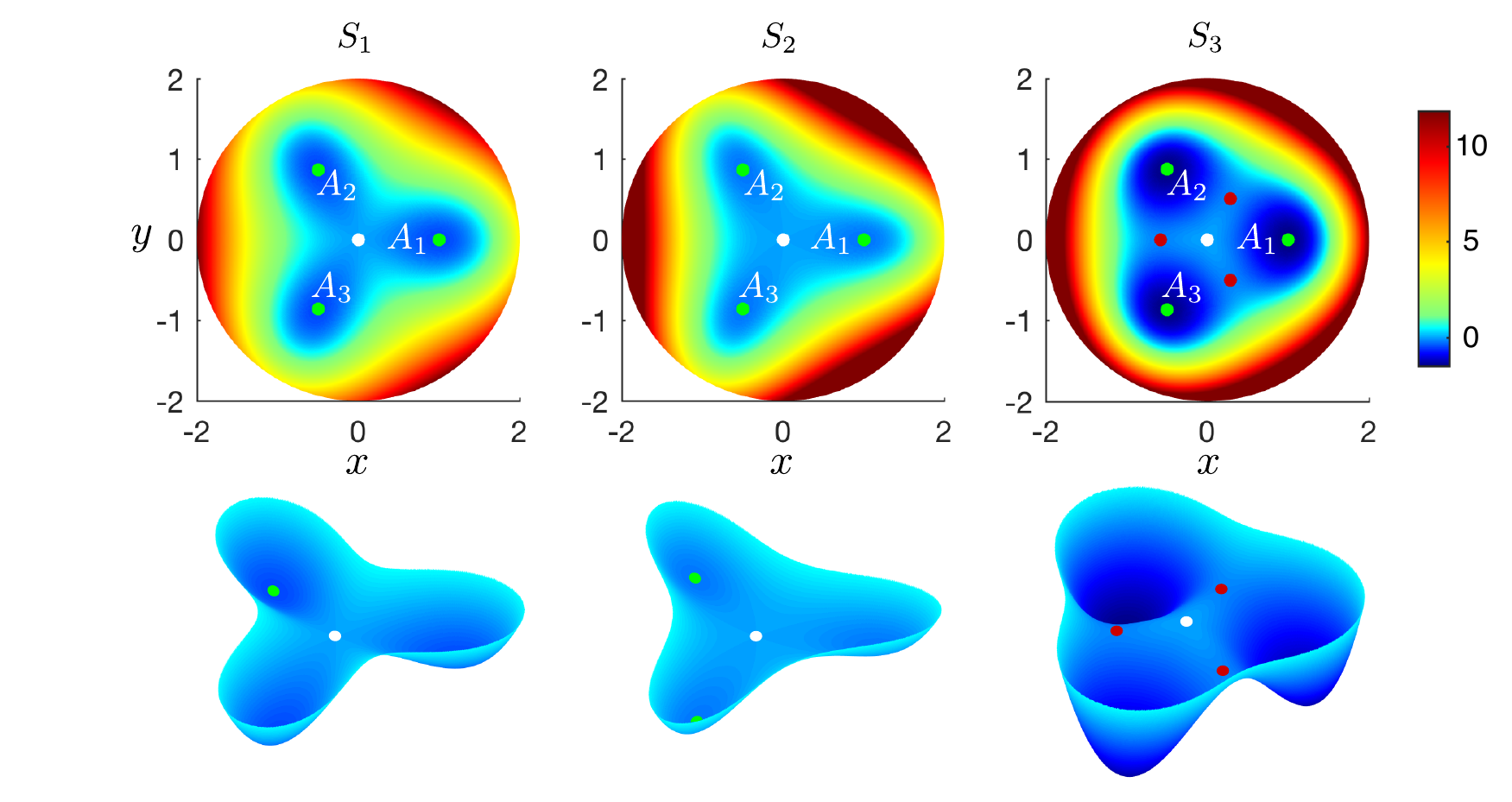}
\end{center}
\vspace{-7mm}
\caption{Three shapes of the roulette surface we consider, given by the functions $S_1$, $S_2$, and $S_3$ defined in the text.  
In each panel, the white dot indicates the unstable fixed point at the origin, the green dots the attractors ($A_1$, $A_2$, and $A_3$), and the red dots the saddle points away from the origin (only present for $S_3$).
The part of each surface corresponding to $S_i(r,\theta)<0.5$ is shown in the bottom row.
Surface colors indicate the value of the function, and a common color scheme is used in all six panels.}
\label{fig:roulette-shapes}
\end{figure*}

As a physical example that can be described using a potential of the form~\eqref{eqn:potential}, consider a roulette system.  When the game is played in reality, a ball is released to a spinning roulette with 38 slots labeled with different numbers. The ball collides multiple times with bumps on the surface of the roulette and eventually falls into one of the slots.  In our study, we simplify this system by assuming that the roulette is still, has a smooth surface, and has three slots (thus $n=3$).  We consider three different shapes of the roulette surface, shown in Fig.~\ref{fig:roulette-shapes} and given by the following functions: 
\begin{align}
S_1(r,\theta) &\equiv - r^2\cos 3\theta + \frac{1}{2}r^4,\\
S_2(r,\theta) &\equiv - r^3\cos 3\theta + \frac{3}{4}r^4,\\
S_3(r,\theta) &\equiv - (2 + \cos 3\theta) r^2 + \frac{3}{2}r^4.  
\end{align}
Note that these functions serve also as the (gravitational) potential of the system, and the three slots correspond to three fixed-point attractors $A_1$, $A_2$, and $A_3$ located at $(r,\theta)=(1,0), (1,\frac{2\pi}{3})$, and $(1,\frac{4\pi}{3})$, respectively.  This means that the results established above apply to this system, implying that the basin boundaries are fractal for $S_1$ [for which $a_2(\theta) = -\cos 3\theta$ takes both positive and negative values], while the boundaries are finite-scale fractal for $S_2$ [for which $a_2(\theta) = 0$] and for $S_3$ [for which $a_2(\theta) = - (2 + \cos 3\theta) < 0$ for all $\theta$, and the surface has three additional saddle points, as indicated by the red dots in Fig.~\ref{fig:roulette-shapes}].  To compensate for the fact that our simplified roulette is still, we consider initial conditions in which the ball is placed on the circle $r=2$ and has a velocity tangent to the circle.  Friction and gravity dominate the motion of the ball.  In order to prevent the ball from moving too far from the center of the roulette, we impose a maximum $v_{\text{max}}(\theta_0)$ on the initial speed $v_0$, where $v_{\text{max}}(\theta_0)$ is defined as the value of $v_0$ corresponding to zero centrifugal acceleration when the ball's initial position is $(2, \theta_0)$ in polar coordinates.  The ball experiences a dragging force proportional to its velocity with coefficient $\mu$ [representing dissipation, as in Eq.~\eqref{eqn:main}], and here we use $\mu = 0.2$.

\begin{figure}[t]
\includegraphics[width=\columnwidth]{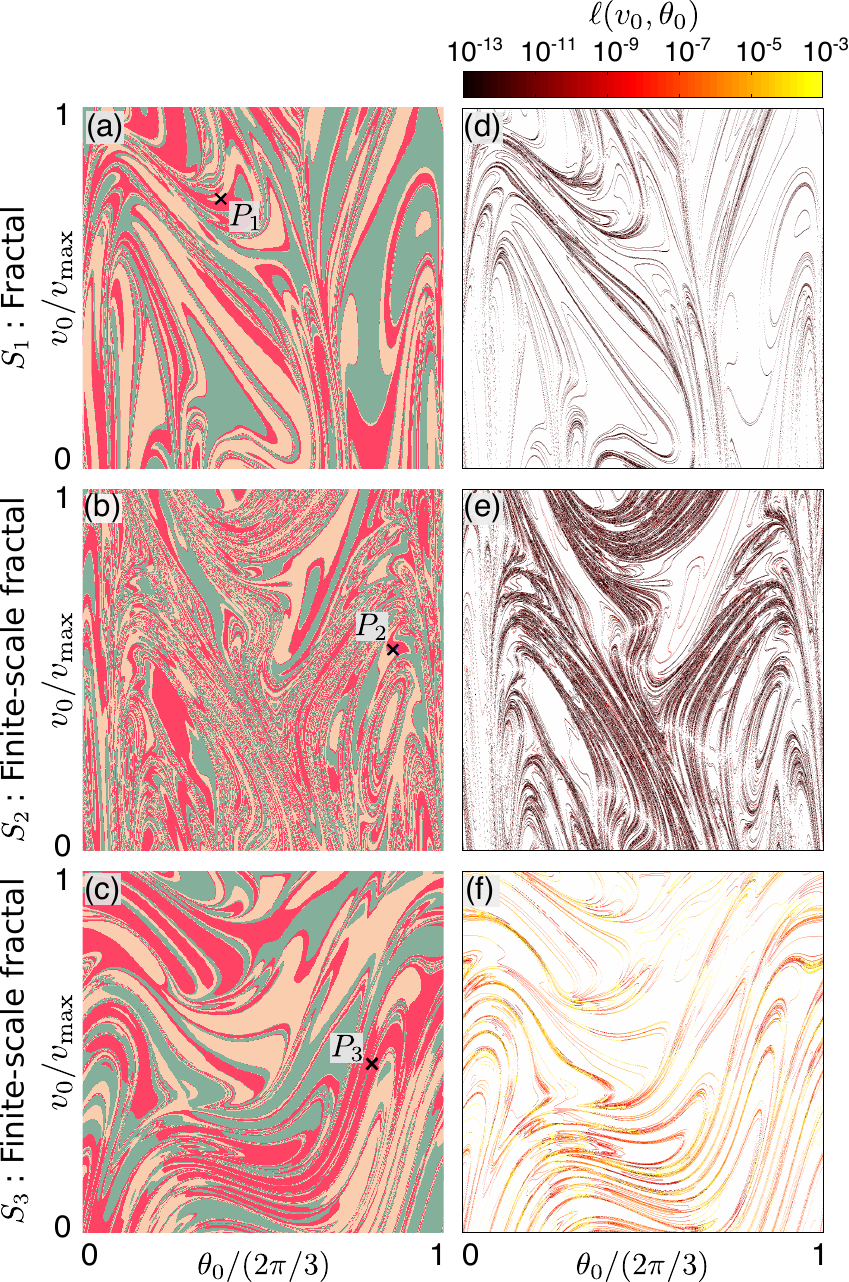}
\vspace{-7mm}
\caption{Slim fractal boundaries of the attraction basins for the roulette system.  (a)--(c) Attraction basins for the roulette shapes shown in Fig.~\ref{fig:roulette-shapes} in the two-dimensional subspace parametrized by the initial condition $(v_0, \theta_0)$.  The red, green, and beige regions indicate the basins of the attractors $A_1$, $A_2$, and $A_3$ (marked in Fig.~\ref{fig:roulette-shapes}), respectively. 
(d)--(f) Spatial distribution of the (color-coded) fractality length scale $\ell(v_0,\theta_0)$ on the boundaries of the basins shown in (a)--(c).
Note that $v_0$ and $\theta_0$ are normalized by $v_{\max}$ and $2\pi/3$, respectively, only for the axes of the plots and not for the computation of $\ell(v_0,\theta_0)$.
We compute $\ell(v_0,\theta_0)$ using double precision and the bisection resolution of $10^{-13}$ for each of the $1{,}024 \times 1{,}024$ grid points [corresponding to $\Delta=2^{-10} \cdot v_{\max}(\theta_0)$, which ranges from $ 3.38 \times 10^{-3}$ to $6.48 \times 10^{-3}$ depending on $\theta_0$ and the roulette shape].}
\label{fig:basin-boundaries}
\vspace{-0.5cm}
\end{figure}

\begin{figure}[t]
\begin{center}
\includegraphics[width=\columnwidth]{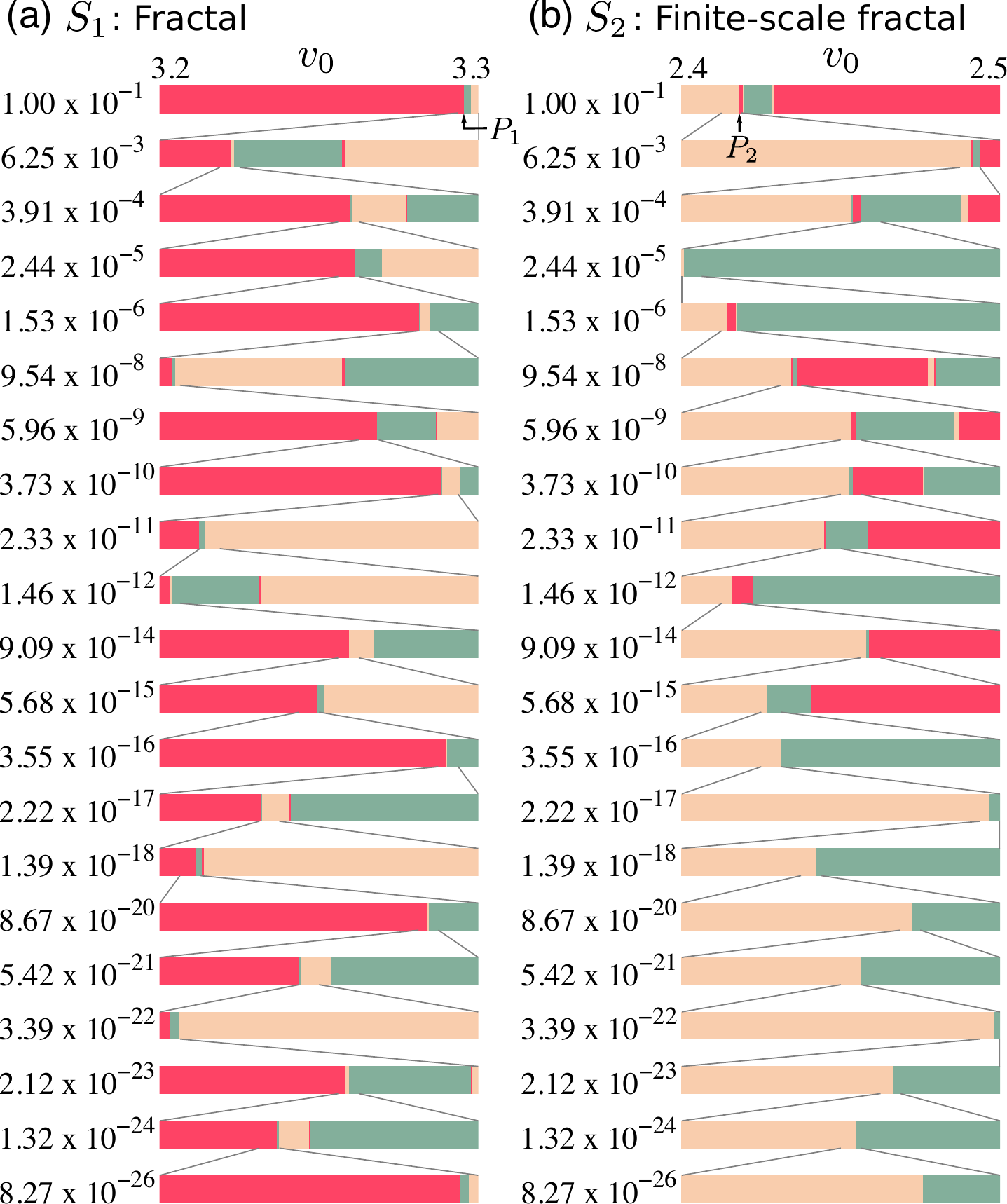}
\vspace{-8mm}
\end{center}
\caption{
Successive magnification of the attraction basins on vertical line segments through the points $P_1$ (a) and $P_2$ (b) in Figs.~\ref{fig:basin-boundaries}(a) and \ref{fig:basin-boundaries}(b), respectively.  
The numbers on the left indicate the length of the magnified intervals.
The details on the computational procedure used to generate this figure can be found in  
Appendix~\ref{sec:succ_mag}.
}
\label{fig:basin_zoom_figure}
\end{figure}

Figures~\ref{fig:basin-boundaries}(a)--\ref{fig:basin-boundaries}(c) show that, despite the difference in the fractality resulting from the three shapes, the numerically estimated boundaries between the basins of the three attractors in the phase space show highly convoluted, fractal-like structures in all three cases.
Comparing Figs.~\ref{fig:basin-boundaries}(a) and \ref{fig:basin-boundaries}(b), we observe that the basin boundaries appear more complex for $S_2$ than for $S_1$.
However, a closer look at the structure around the points $P_1$ and $P_2$ [marked in Figs.~\ref{fig:basin-boundaries}(a) and \ref{fig:basin-boundaries}(b), respectively] through successive magnifications of one-dimensional cross sections in Fig.~\ref{fig:basin_zoom_figure} reveals a surprise: while for $S_1$ a new basin keeps appearing upon magnification [Fig.~\ref{fig:basin_zoom_figure}(a)] even at the limit of numerical precision used for integration (on the order of $10^{-31}$), the magnification plots for $S_2$ [Fig.~\ref{fig:basin_zoom_figure}(b)] show that the basin boundary is simple below a certain finite length scale (on the order of $10^{-15}$).  To systematically quantify this \textit{fractality length scale}, consider applying the bisection algorithm to a small vertical line segment of length $\Delta$ in the $(v_0, \theta_0)$-space, which can be used to estimate the location of a boundary point (to a given numerical resolution).  We define $\ell(v_0, \theta_0)$ to be the length of the interval used in the last occurrence of the following situation in the bisection process: the midpoint belongs to a basin that differs from those to which the two end points belong.  
For example, the quadruple-precision bisection procedure used to generate the magnification plots in Fig.~\ref{fig:basin_zoom_figure} for $P_1$ and $P_2$ gives $\ell \approx 2.58 \times 10^{-27}$ and $\ell = 1.42 \times 10^{-15}$, respectively (with $\Delta = 0.1$ and resolution on the order of $10^{-27}$, see 
Appendix~\ref{sec:succ_mag}
for details). 
Note that the fractality length scale at $P_2$ is at the limit of double-precision calculation and thus could not be clearly resolved without using higher precision.
This illustrates the fact that a finite-scale fractal can be numerically indistinguishable from true fractals.
The fractality length scale can also be seen as a quantitative measure of the Wada property at a given point (see Ref.~\cite{Daza2014} for a different numerical approach to quantify this property).

The fractality length scale $\ell(v_0, \theta_0)$ can generally depend on phase-space location $(v_0, \theta_0)$, and its spatial distribution is quite different for the three example shapes [see Figs.~\ref{fig:basin-boundaries}(d)--\ref{fig:basin-boundaries}(f)]. 
For $S_1$, the computed length scale $\ell$ is at the chosen precision ($=10^{-13}$) uniformly over the boundary set (although the exact number depends slightly on the details of each bisection sequence), which is consistent with the true fractality of the boundaries.
For both $S_2$ and $S_3$, the boundaries are finite-scale fractals, and for $S_3$ the length scale $\ell$ is indeed well above the scale of the chosen precision across the boundary set.  In contrast, $\ell$ shows a mixed behavior for $S_2$, where $\ell$ is close to the scale of the chosen precision for the most part, but is well above that scale in certain locations.  In this sense, the finite-scale fractal for $S_2$ is closer to a true fractal than for $S_3$.  Further analysis of the probability distribution of $\ell$, as well as of a quantitative measure of the Wada property, corroborates these observations (see 
Appendix~\ref{sec:dist_frac_meas}).

We expect to see similar geometry of the basin boundaries if we consider the more realistic case of a roulette rotating at a constant angular velocity with zero initial velocity for the ball.  
Rewriting Eq.~\eqref{eqn:main} in the frame co-rotating with the roulette,  we gain two additional terms representing the centrifugal and Coriolis forces.  
The former effectively adds a constant to the coefficient $a_2(\theta)$ in Eq.~\eqref{eqn:potential}, while the latter simply shifts the location of the basin boundaries without altering the fractality of the boundaries.

\section{Equivalent dimension for slim fractals}

\begin{figure*}[tb]
\begin{center}
\includegraphics[width=6in]{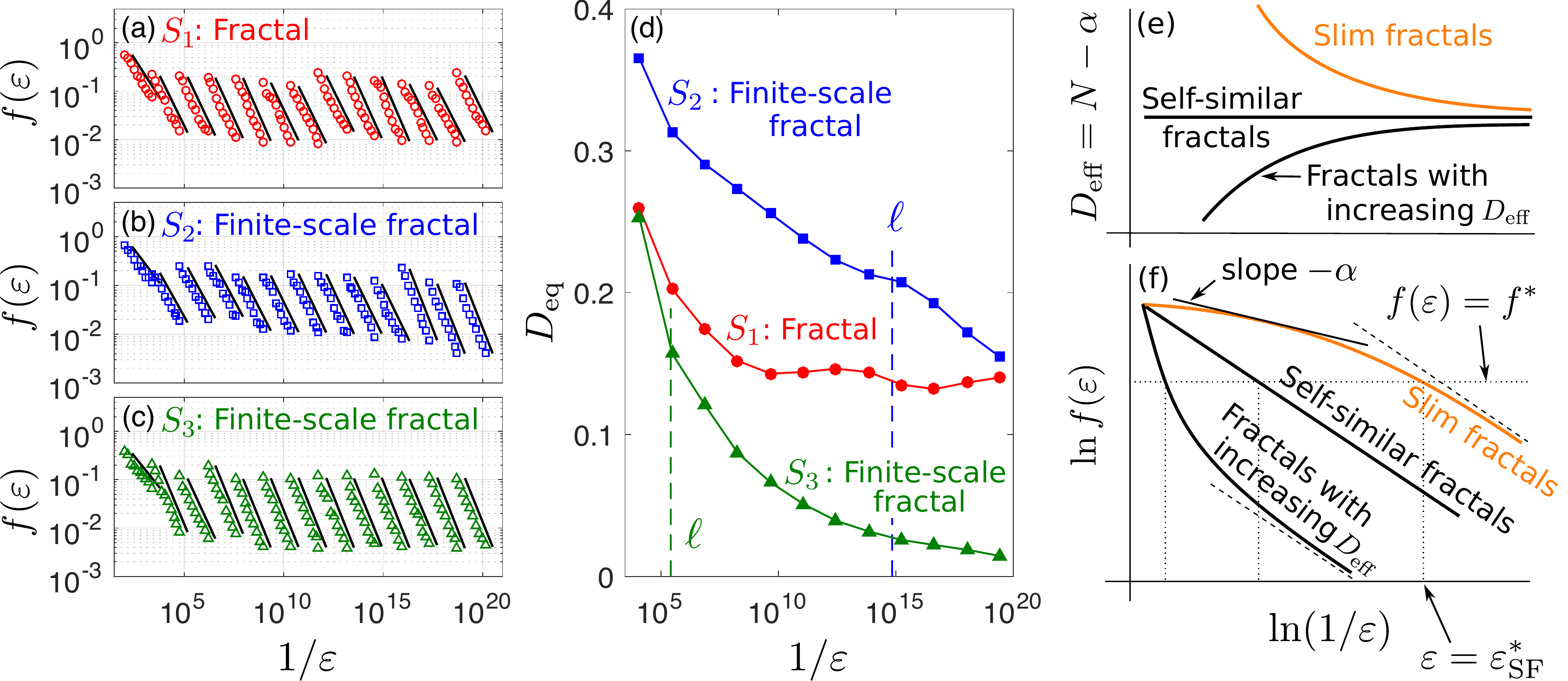}
\vspace{-5mm}
\end{center}
\caption{Dimension of slim fractals.
(a)--(c) Uncertainty function $f(\varepsilon)$ estimated by sampling pairs of points $\varepsilon$ apart on a vertical line segment centered at $P_1$, $P_2$, and $P_3$ in Fig.~\ref{fig:basin-boundaries} for the roulette system with surface shapes $S_1$, $S_2$, and $S_3$, respectively, for successively smaller segments (left to right).
(d) Equivalent dimension $D_\text{eq}$, computed using Eq.~\eqref{eq:D-eq} with $L=2\times10^{-3}$ and $D_\text{eff}$ estimated as the slope for each segment in (a)--(c) from the linear least-squares fit [lines in (a)--(c) are offset for clarity]. The vertical dashed lines indicate the fractality length scale $\ell$ for $S_2$ and $S_3$.
(e),(f) Illustration of the effective dimension (e) and the final-state uncertainty (f) as a function of $\ln(1/\varepsilon)$ for three types of fractals.}
\label{fig:eff_dim_fig}
\end{figure*}

The fractality of the basin boundaries can be quantitatively characterized also by their dimension, which can be defined through a scaling relation between initial-state accuracy and final-state uncertainty~\cite{ott,ua}.
For a self-similar system and an $N$-dimensional region of its phase space, the scaling is $f(\varepsilon)\sim \varepsilon^{N-D}$, where the constant $D$ is defined as the fractal dimension of the boundaries, and $f(\varepsilon)$ is the final-state uncertainty, defined as the fraction of pairs of points belonging to different basins among all pairs that are within the region and $\epsilon$ apart from each other.
In contrast, the scaling exponent is resolution dependent for the systems studied here [as shown in Figs.~\ref{fig:eff_dim_fig}(a)--\ref{fig:eff_dim_fig}(c) for the roulette system], which motivates us to adopt a finite-scale measure of the dimension.
With that in mind, we first consider  using the effective fractal dimension~\cite{Motter:2003,Karolyi:2005,motter05} given by
\begin{eqnarray}\label{eq:effdim}
D_\text{eff}(\varepsilon)&=&N - \frac{d\ln f(\varepsilon')}{d\ln\varepsilon'}\Bigr|_{\varepsilon'=\varepsilon},
\end{eqnarray}
which is a strictly local measure of how uncertainty changes with resolution.
Specifically, the effective dimension describes the relation between small improvement in initial-state accuracy and the resulting reduction in final-state uncertainty at the finite scale $\varepsilon$. 
The usual (asymptotic) definition of fractal dimension is recovered in the limit $\varepsilon \rightarrow 0$.

In general, for {\it slim fractals}---which we define as having $D_\text{eff}$ that decreases with decreasing $\varepsilon$---the effective dimension at a given scale fails to capture the complexity of the basin boundaries observed at larger scales and its impact on the dynamics. 
To see this, consider the case of finite-scale fractals, for which we have $D_\text{eff} = N-1$ below the fractality length scale $\delta>0$.
In this case, the final-state uncertainty scales as $f(\varepsilon) \sim \varepsilon^{N-D} = \varepsilon$, which is the same as that of a system without sensitive dependence on initial conditions. 
This means that the improvement in the accuracy of initial conditions (i.e., the amount by which $\varepsilon$ is reduced) required to achieve a given level of uncertainty can be much less compared to the case of fractal boundaries with $N-D < 1$.
However, a prerequisite for benefiting from this linear scaling is that $\varepsilon < \delta$, which is itself a requirement on the accuracy of initial conditions. 
A similar argument applies to the case of true (but slim) fractals, since benefiting from smaller $D_\text{eff}$ (thus larger scaling exponents) requires the initial condition accuracy to be high in the first place.

To characterize the finite-scale sensitive dependence on initial conditions, we define a new dimension $D_{\text{eq}}(\varepsilon)$ to be the dimension of an equivalent self-similar system, whose final-state uncertainty is the same as the system being studied at two different scales: $\varepsilon$ and a larger reference scale $L$.
We term this quantity {\it equivalent dimension} and show that it can be expressed as
\begin{eqnarray}\label{eq:D-eq}
D_\text{eq}(\varepsilon)&=&\frac{1}{\ln L - \ln\varepsilon}\int_{\varepsilon}^L\frac{D_\text{eff}(\varepsilon')}{\varepsilon'} d\varepsilon' \mbox{,}
\end{eqnarray}
which, as an integral quantity, properly accounts for the cumulative impact of the effective dimension on the relation between initial-state accuracy and final-state uncertainty in the systems we consider.
The equivalent dimension in Eq.~\eqref{eq:D-eq} can be derived as follows.
First, writing the final-state uncertainty of the equivalent self-similar system as $\tilde{f}(\varepsilon') = C\cdot(\varepsilon')^{N-D_\text{eq}}$, where $C$ is a constant, we have $f(L) = CL^{N-D_\text{eq}}$ and $f(\varepsilon) = C\varepsilon^{N-D_\text{eq}}$.
Next, we eliminate $C$ from these two equations and obtain $D_\text{eq}=N-\frac{\ln f(L) - \ln f(\varepsilon)}{\ln L - \ln\varepsilon}$.
Since this can also be obtained by using Eq.~\eqref{eq:effdim} and rewriting Eq.~\eqref{eq:D-eq}, we see that the equivalent dimension is indeed given by Eq.~\eqref{eq:D-eq}.
Thus, we have a more intuitive and direct definition of fractal dimension that considers the entire process of decreasing $\varepsilon$ to improve the accuracy of predicting the final state.

For the case of finite-scale fractals, which have fractal dimension $D=N-1$, the dependence of the equivalent dimension on $\varepsilon$ is given by the general formula
\begin{equation}\label{eq:D-eq-theory}
D_\text{eq}(\varepsilon) =
D + (D_\text{eq}(\delta)-D)\cdot\frac{\ln L - \ln\delta}{\ln L - \ln\varepsilon}
\end{equation}
for $\varepsilon < \delta$ [which follows directly from Eqs.~\eqref{eq:effdim} and \eqref{eq:D-eq}].
When $D_\text{eq}(\delta) > D$, we see that $D_\text{eq}(\varepsilon)$ slowly (and continuously) decreases from $D_\text{eq}(\delta)$  and approaches $D$ as $\varepsilon\to0$.
Thus, the equivalent dimension for scales below $\delta$ ``feels'' the effect of large $D_\text{eq}(\delta)$ (and hence of $D_\text{eff}$ at scales larger than $\delta$), which reflects the sensitivity to initial conditions observed at scales above $\delta$.  
While we do not expect Eq.~\eqref{eq:D-eq-theory} to be followed exactly in practice, as the scaling of $f(\varepsilon)$ is never perfect, we do expect $D_\text{eq}(\varepsilon)$ to start decreasing at the fractality length scale and approach the asymptotic dimension $D=N-1$.
This is indeed observed in Fig.~\ref{fig:eff_dim_fig}(d) for one-dimensional cross sections (thus $D=N-1=0$) of the basin boundaries in our roulette system with shapes $S_2$ and $S_3$.
For $S_1$, with the basin boundaries forming a true fractal, the equivalent dimension seems to approach $D_\text{eq}\approx 0.14$.
The uncertainty-based calculations for all three cases are consistent with the results from another numerical approach (valid for $N=1$), based on the fractal dimension estimate,
\begin{equation}\label{eqn:dim-interval-len}
D = - \frac{\ln 2}{\displaystyle\lim_{i \to \infty}\ln(\ell_{i+1}/\ell_i)},
\end{equation}
where $\ell_i$ is the length of the $i$th interval identified as part of the Cantor set structure of the basin boundaries (see 
Appendix~\ref{sec:basin_int_length}
for details, where we account for intervals as small as $\ell_i=1.1 \times 10^{-27}$).
Interestingly, Fig.~\ref{fig:eff_dim_fig}(d) shows that the equivalent dimension of the finite-scale fractal for $S_2$ is significantly larger than that of the true fractal for $S_1$ for scales above $10^{-20}$.
This, however, is actually consistent with the more complex basin boundaries observed for $S_2$ [Fig.~\ref{fig:basin-boundaries}(b)] than for $S_1$ [Fig.~\ref{fig:basin-boundaries}(a)].

The equivalent dimension fills a gap between classes of systems that can be suitably characterized with existing definitions.
For self-similar systems, $D_\text{eff}$ is constant, as illustrated in Fig.~\ref{fig:eff_dim_fig}(e), which corresponds to a straight line for the graph of $\ln f(\varepsilon)$ vs.\ $\ln(1/\varepsilon)$, as illustrated in Fig.~\ref{fig:eff_dim_fig}(f).
In this case, the complexity of the basin boundaries is captured well by the usual asymptotic definition of fractal dimension $D$ (and by $D_\text{eff}$ at any finite $\varepsilon$).
For non-hyperbolic systems (such as Hamiltonian systems with mixed phase space~\cite{motter05}), $D_\text{eff}$ increases as a function of $\ln(1/\varepsilon)$~\cite{Lau:1991}, as shown in Fig.~\ref{fig:eff_dim_fig}(e), and this corresponds to a convex curve in Fig.~\ref{fig:eff_dim_fig}(f).
In this case, the asymptotic dimension $D$ reflects the complex geometry of the basin boundaries [and is lower bounded by $D_\text{eff}(\varepsilon)$ for finite $\varepsilon$].
In contrast, in the class of undriven dissipative systems we consider here, $D_\text{eff}$ decreases as a function of $\ln(1/\varepsilon)$, as shown in Fig.~\ref{fig:eff_dim_fig}(e) [which is directly associated with the decrease of $D_\text{eq}$ as a function of $1/\varepsilon$ observed in Fig.~\ref{fig:eff_dim_fig}(d)], and this corresponds to the concave curve in Fig.~\ref{fig:eff_dim_fig}(f). 
This behavior of $D_\text{eff}$ is the defining characteristic of slim fractals and reflects their structure, which appears sparser at smaller length scales.
Since $D_\text{eff}(\varepsilon)\ge D$ in this case, $D$ is only a ``lower bound'' for the finite-scale geometrical complexity reflected in $D_\text{eff}(\varepsilon)$, and can in fact indicate no complexity at all (e.g., the case of finite-scale fractals with asymptotic dimension $D=N-1$, which equals the dimension of simple boundaries).
Figure~\ref{fig:eff_dim_fig}(f) illustrates that the shape of the graph of $\ln f(\varepsilon)$ vs.\ $\ln(1/\varepsilon)$ determines the initial condition accuracy required to achieve a given level of uncertainty $f(\varepsilon)=f^*$.
The concavity of this graph for slim fractals implies that the required initial condition accuracy $\varepsilon^*_\text{SF}$ can be orders of magnitude smaller than the corresponding numbers for the other types of fractals, even when the asymptotic dimension [and thus the asymptotic slope of the curves in Fig.~\ref{fig:eff_dim_fig}(f)] is the same.
By design, $D_\text{eq}$ integrates the finite-scale complexity over a range of different scales, and is therefore suitable for studying such systems.
As an integral of $D_\text{eff}$, the equivalent dimension also enjoys the benefit of having less numerical errors than $D_\text{eff}$.

\section{Discussion}

We have demonstrated that the basin boundaries in systems exhibiting doubly transient chaos are generically true fractals, with both Cantor set structure and the Wada property observed at arbitrarily small length scales.
It is instructive to compare this with the most previously studied forms of transient chaos (i.e., those in driven or conservative systems). 
In all cases, the basin boundaries correspond to the stable manifolds of an unstable invariant set. However, this set consists of an uncountable number of trajectories in previous cases but of only one unstable fixed point (the origin) in the systems  considered here. 
Accordingly, the basin boundaries consist of one or a few manifolds in our case, as opposed to a bundle of uncountably many manifolds as in previously studied cases. 
But can a finite number of manifolds really define a fractal?
The answer has long been known to be yes; the Koch snowflake is an immediate example---though the curve is non-differentiable and constructed {\it ad hoc}---but there are also known examples of a dynamically generated manifold forming a fractal, such as the invariant manifolds in homoclinic tangles~\cite{Guckenheimer:2013}.
Therefore, our result that such boundaries are true fractals is not the first demonstration of fractal geometry arising from a finite number of manifolds.
However, an interesting aspect of the fundamental problem studied here is that, contrary to the case of homoclinic tangles, which embed Smale horseshoes with (permanent) chaotic trajectories, our dissipative systems cannot exhibit any sustained oscillations (chaotic or otherwise): every system trajectory must converge to an equilibrium.
This underlies the fact that the stable manifold of a single equilibrium is fully responsible for the complexity of the fractal basin boundaries in the systems we consider.

We have also demonstrated that, even when the boundaries do not form a true fractal, they can give rise to a form of sensitive dependence on initial conditions, 
which nevertheless is not properly characterized by existing notions of dimension.
These results challenge us to think differently about the definition of fractals. In many natural systems, geometric structures similar to fractals are observed, but they disappear at sufficiently small scales due to finite resolution or the nature of physics at that length scale. 
Nonetheless, those systems are likely to exhibit sensitive dependence on initial conditions at physically relevant length scales (e.g., those relevant for measuring the initial state).
For example, games of chance, such as a dice roll, are undriven dissipative systems for which the basin boundaries can be simple at sufficiently high resolution~\cite{dyn-gambling-book,Sato2015}, but there are no practical methods to measure initial conditions at that resolution and predict the outcomes.
Moreover, our results show that the resolution below which boundaries become simple can be highly dependent on the phase-space location.
An immediate option for studying such systems is to use an existing notion of scale-dependent dimension, such as the effective dimension.
However, for being a local measure of uncertainty versus length scale, the effective dimension alone cannot capture the physically observable sensitive dependence on initial conditions.
Our integral-based definition of the equivalent dimension addresses this issue and, together with the fractality length scale, offers an analysis framework for studying undriven dissipative systems.

Our findings have profound implications for the physics of undriven dissipative systems.
Prominent examples include the following:

\begin{enumerate}[leftmargin=0.17in]

\item {\it Astrophysical systems.}
When two compact objects---e.g., neutron stars, white dwarf stars, or black holes---orbit each other emitting gravitational waves, we have an undriven dissipative system (since energy is lost due to gravitational radiation)~\cite{Hartl:2003}.
Such coalescing binary systems serve as candidate sources of detectable gravitational waves.
Characterizing the dynamics and geometry of these systems has been controversial, with arguments both for~\cite{Levin:2000} and against~\cite{Cornish:2000} the existence of chaos and fractal basin boundaries.
This issue is significant because sensitive dependence on initial conditions would lead to an explosion in the number of possible theoretical templates of gravitational waves against which the observational data would have to be matched, necessitating alternative detection methods.

\item {\it Fluid systems.}
Interacting vortices in an otherwise still viscous fluid form undriven dissipative systems whose characterization of chaos is relevant and to which existing tools do not apply.
Typically, scenarios involving three or more vortices are considered to allow for chaotic dynamics.
In part because of the lack of adequate tools, previous studies of chaotic dynamics in such systems focused primarily on potential flows and other solutions of the Euler equations (in which dissipation due to viscosity is neglected)~\cite{Aref:1983}. 
Our results established here open the possibility of a self-consistent study of chaos in solutions that properly account for viscous dissipation. 

\item {\it Chemical systems.}
Nonlinear chemical reactions in thermodynamically closed systems can exhibit chaotic dynamics in the absence of any driving~\cite{Scott:1991,Yatsimirskii:1993}.
Previous studies of such systems, of which the Belousov-Zhabotinsky reaction is an example, have focused primarily on the far-from-equilibrium regime of strong chaotic oscillations.
This regime is nevertheless transient, as dissipation unavoidably makes the system approach thermodynamic equilibrium.
Our results can allow the complete characterization of this transition to equilibrium, which thus far could be only partially understood using the tools of conservative and driven dissipative systems.

\end{enumerate}

Ultimately we note that our derivation of the fractality condition and the measures introduced here to quantify slim fractals
do not rely on the specifics of the systems considered.
Thus, we expect these results to be generalizable to undriven dissipative systems exhibiting doubly transient chaos in higher dimensions and with an arbitrary number of basins.

\begin{acknowledgments}
The authors thank Gy\"orgy K\'arolyi and Yuzuru Sato for insightful discussions.
This work was supported by a  Krieghbaum Scholarship, a WCAS Summer Grant, and the NSF Grant under No.\ PHY-1001198.
\end{acknowledgments}

\appendix

\section{Derivation of the fractality condition}
\label{sec:full-analysis}

\noindent{\it Fractal for potential $U_2(r,\theta)$}.
Consider the set of trajectories with initial velocity zero and initial positions on the vertical line segment labeled A in Fig.~\ref{fig:potential_syst}(a).  As illustrated by a few representative trajectories, the green, red, and orange parts of the segment belong to the basins of the exits $E_1$, $E_2$, and $E_3$, respectively.  Now consider the trajectories initiated near the gap between the top green part and the middle red part of the segment.  These trajectories climb the hill and turn around when they reach the curved segment labeled B, after which some trajectories move back toward the origin.  Such trajectories will approximately trace a trajectory on the line $\theta = \pi/3$, $r\ge 0$, which is governed by the equation, $\ddot{r} + \mu\dot{r} + 2r = 0$.  Solving this equation and assuming that the dynamics is under-damped ($\mu < \mu_c$, where $\mu_c \equiv 2\sqrt{2}$), we see that this trajectory reaches the origin in finite time with {\it nonzero} velocity.  This implies that the trajectory that turned around and approached the neighborhood of the origin continues moving and exits through $E_3$.  
Thus, there is a small orange segment within the gap between the green and red parts of line segment A.  This argument can be repeated for the trajectories leaving from the gap between the red and orange portions of segment B, and we find that some of these trajectories are deflected one additional time when they reach segment C, and then exit through $E_2$.  
This implies that there is an even smaller green segment of A in the gap between the orange and red portions.  The argument can be repeated indefinitely, which shows that, between any two segments of different colors on A, one can always find a segment of the third color.  

Note that the qualitative argument above is valid in general as long as all trajectories moving downhill toward the origin eventually leave the scattering region through the exit on the other side.
It can also be extended to cases in which segment A is replaced by an arbitrary line segment connecting two points from different basins in the full phase space.
The trajectories from these two points generally bounce between the three hills some number of times before exiting the scattering region.
If the patterns of these bounces for the two trajectories are different, one can slide (along the line segment) the initial point of the trajectory with fewer bounces closer to the other point until the patterns of bounces match (while ensuring that they still belong to different basins).
We can then apply the same argument as before to see that, between these two initial points (which belong to two different basins), there must be another initial point that belongs to the third basin (with its trajectory approaching the origin and then leaving through the third exit).

\medskip

\noindent{\it Finite-scale fractal for potential $U_3(r,\theta)$}.
In this case, the dynamics on the line $\theta = \pi/3$, $r\ge 0$, which is governed by $\ddot{r} + \mu\dot{r} + 3r^2 = 0$, is effectively over-damped for arbitrary $\mu>0$ when the trajectory is sufficiently close to the origin.
This underlies the existence of the finite line segment $\{-r_s \le x \le 0,\,\, y=0\}$ from which all trajectories approach the origin asymptotically, which supports the argument in the main text leading to the non-Wada property of these boundary points.
To show that this segment splits into two branches forming simple boundaries, consider a vertical segment at $x < -r_s$, such as segment B$'$ at $x=-0.5$ in Fig.~\ref{fig:potential_syst}(c). Because $x < -r_s$, there is a part of this segment from which trajectories eventually exit through $E_1$, but the boundaries between different basins are simple.  This is due to the presence of a trajectory that is deflected by the hill at $\theta=\pi/3$ before approaching the origin as $t\to\infty$ (black curve), similarly to the one starting on segment A$'$ and approaching the origin asymptotically.  The same argument as above applied to this trajectory shows that the boundary between basins of $E_1$ and $E_3$ is simple. Thus, the simple segment of the boundary touching the origin splits into two branches forming simple boundaries (the blue arrow indicates the branching point at $x = - r_s$, $y = 0$).  Repeating this argument with segment C$'$ and other similar segments of initial positions, we see that the basin boundaries form a binary tree of simple segments.
We observe that, as one moves away from the origin along the branching tree, the gaps between branches narrow, thus making the fractality length scale smaller.
Since we have assumed zero initial velocities, the binary tree we just established is a two-dimensional cross section of the basin boundaries in the full four-dimensional phase space.

To see that this full set of boundaries is also not truly fractal, 
we apply the center manifold reduction~\cite{Guckenheimer:2013} to the equilibrium at the origin.
The Jacobian matrix at the origin has eigenvalues $0$ and $-\mu$, each with multiplicity $2$.
This implies that there exists a two-dimensional center manifold and a two-dimensional stable manifold in a neighborhood of the origin.
Writing Eq.~\eqref{eqn:main} in terms of the eigenvector coordinates 
$(\tilde{x},\tilde{y},\tilde{u},\tilde{v}) \equiv (x+\dot{x}/\mu, y+\dot{y}/\mu, -\dot{x}/\mu, -\dot{y}/\mu)$, we determine the center manifold and the dynamics on it to be given by Eqs.~\eqref{eq:center_manifold} and \eqref{eq:center_manifold_dyn}, respectively, up to second order in $\tilde{x}$ and $\tilde{y}$.
Figure~\ref{u_3_center_manifold_fig}(b) shows that the region is divided into three basins (corresponding to exits $E_1$, $E_2$, and $E_3$) by three segments of simple boundaries: the half lines $\theta=\pi/3$, $\theta=\pi$, and $\theta=5\pi/3$.
Since the stable manifold is two dimensional, these boundaries on the center manifold extend to three pieces of simple, smooth, and thus non-fractal three-dimensional boundaries dividing a four-dimensional neighborhood of the origin.
These boundaries, when extended as much as possible, intersect with the subspace $\dot{x} = \dot{y} = 0$ in the line segments $0 \le r \le r_s$, $\theta=\pi/3$, $\pi$, $5\pi/3$, in Fig.~\ref{fig:potential_syst}(c).
The full set of basin boundaries can then be expressed as the set of all points whose trajectory ultimately falls on one of these local boundaries.
This is because approaching the origin is the only asymptotic behavior possible for the system besides leaving the scattering region.
Thus, in a sufficiently small neighborhood of any basin boundary point, the boundary is a three-dimensional smooth manifold, since it is a pre-image of part of the local boundaries near the origin.
Therefore, the global basin boundaries, whose two-dimensional cross section is the binary tree we established above, are not fractal but form a finite-scale fractal inheriting the branching structure.

\section{Class of potentials $U_{\alpha}$ with arbitrary $\alpha$}
\label{sec:U-alpha}

We show that the arguments in Appendix~\ref{sec:full-analysis} are also valid for the class of potential functions $U_\alpha(r,\theta) \equiv -r^\alpha\cos(3\theta)$, which then implies that the basin boundaries form a fractal for $\alpha\le 2$ and a finite-scale fractal for $\alpha> 2$.  In other words, as $\alpha$ decreases through the transition point $\alpha=2$, the basin boundaries transform from a branching tree structure to a shape similar to a Cantor fan that exhibits fine basin structures at any resolution.  To see why $\alpha=2$ is the transition point between the two regimes, note that we can write $U_\alpha(r,\theta) = -\beta(r)r^2\cos(3\theta)$, where $\beta(r) \equiv r^{\alpha - 2}$ can be interpreted as an $r$-dependent prefactor for the quadratic potential $U_2$. According to this interpretation, the dynamics on line $\theta = \pi/3$, $r\ge 0$ [governed by $\ddot{r} + \mu\dot{r} + 2\beta(r)r = 0$] would be critically damped if $\xi \equiv \frac{\mu}{2\sqrt{2\beta(r)}} = 1$, over-damped if $\xi > 1$, and under-damped if $\xi < 1$.  For $\alpha > 2$, the same argument we used in the main text for the case $\alpha=3$ can be used to show that the basin boundaries form a finite-scale fractal, since the dynamics is also effectively over-damped in this case as long as $r < \bigl(\frac{\mu^2}{8}\bigr)^{\frac{1}{\alpha-2}}$.  In comparison, for $\alpha \le 2$, there is a neighborhood of the origin in which the dynamics is effectively under-damped.  We can thus use the same argument used above for the case $\alpha=2$ to establish the fractality of the basin boundaries.
This fractality transition at $\alpha=2$ is numerically verified in Appendix~\ref{sec:frac_trans_U-alpha}.
In the more general case of nonlinear $\mu=\mu(r,\theta) \ge0$, the condition for the boundaries to be a finite-scale fractal (true fractal) is that $\xi(r) = \frac{\mu(r,\,\pi/3)}{2\sqrt{2\beta(r)}} > 1$ ($<1$) for all $r$ sufficiently small.
If the dissipation is of the form $\mu(r,\theta)=\mu_0 r^q$, $q>0$, for example, the fractality transition occurs at $\alpha=2(1+q)$.

\section{Fractality transition for $U_{\alpha}$}
\label{sec:frac_trans_U-alpha}

\begin{figure}[t]
\begin{center}
\includegraphics[width=\columnwidth]{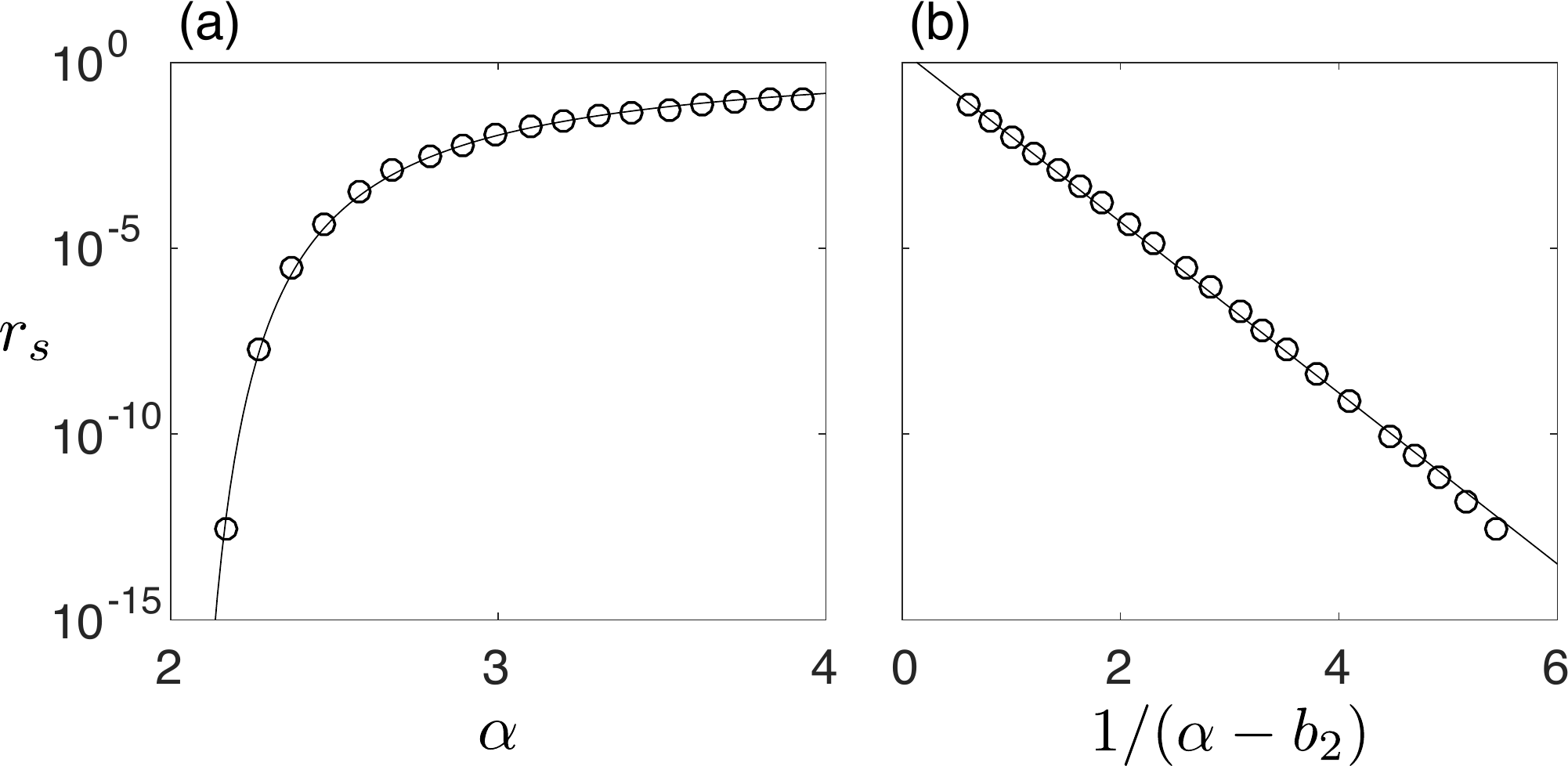}
\vspace{-9mm}
\end{center}
\caption{Length $r_s$ of the simple boundary segment on the line $y=0$ for systems with potential $U_{\alpha}$.
(a) The circles indicate the numerically estimated $r_s$ as a function of $\alpha$.  
The solid curve is the best nonlinear fit to the data, as described in Appendix~\ref{sec:frac_trans_U-alpha}.
(b) The same quantities shown as a function of $1/(\alpha-b_2)$, with the best-fit value of $b_2 \approx 1.986$.}
\label{fig:rsp_se}
\end{figure}

To numerically verify that the transition occurs at $\alpha=2$, we estimate the length $r_s$ of the simple boundary segment $\{-r_s \le x \le 0,\,\, y=0\}$ and confirm that this length approaches zero as $\alpha \to 2$ from above.  
As explained in Appendix~\ref{sec:full-analysis}, all trajectories starting on this segment approach the origin asymptotically, while any trajectory starting on the line segment $\{x<-r_s,\,\, y=0\}$ passes through the origin and exits through $E_1$.  We thus compute $r_s$ by locating the point $(-r_s,0)$ using the following two-level bisection method:
\begin{enumerate}
\item For a given $x$, use the bisection algorithm to determine the top and bottom boundary points of the basin of $E_1$ on the vertical line at $x$.
\item Apply the bisection method on the $x$ value to determine $x = -r_s$ as the boundary point between those $x$ values for which the basin of $E_1$ is found and those for which the basin is not found.
\end{enumerate}
As shown in Fig.~\ref{fig:rsp_se}(a), the numerically estimated length $r_s$ decreases to zero as $\alpha$ approaches two, but reaches the machine (double) precision ($\approx 10^{-15}$) well above $\alpha=2$.  The condition for over-damping mentioned in Appendix~\ref{sec:U-alpha} suggests the relation $r_s \sim \bigl(\frac{\mu^2}{8}\bigr)^{\frac{1}{\alpha-2}}$.  We thus fit the nonlinear function $f(\alpha; b_1, b_2) = b_1\bigl(\frac{\mu^2}{8}\bigr)^{\frac{1}{\alpha-b_2}}$ with two tunable parameters $b_1$ and $b_2$ to the estimated values of $r_s$.  
The least-squares fit for $\ln r_s$, shown in Fig.~\ref{fig:rsp_se}(b), yields $b_1 \approx 2.055$ and $b_2 \approx 1.986$, which is consistent with our claim, $b_2 = 2$.  
We thus have numerical evidence that the fractality transition takes place at $\alpha = 2$.  

\section{Computational procedure for Fig.~\ref{fig:basin_zoom_figure}}
\label{sec:succ_mag} 

Figure~\ref{fig:basin_zoom_figure} is generated by applying the bisection algorithm to the vertical line segment $3.2 \le v_0 \le 3.3$, $\theta_0 = 0.8$ (passing through $P_1$ at $v_0 \approx 3.295$) and the segment $2.4 \le v_0 \le 2.5$, $\theta_0 = 1.8$ (passing through $P_2$ at $v_0 \approx 2.418$).  
Both intervals thus have length $\Delta=0.1$. In each iteration, we determine the basin to which the midpoint of the interval belongs by integrating the system with quadruple precision and relative accuracy of $10^{-4}$ (with respect to the length of the bisection interval for that iteration).
We iterate until the interval length becomes equal to $2^{-86} \cdot \Delta \approx 1.29 \times 10^{-27}$ and $2^{-84} \cdot \Delta \approx 5.17 \times 10^{-27}$ for $P_1$ and $P_2$, respectively.  Numerical integration on these intervals is thus performed with absolute accuracy of $1.29 \times 10^{-31}$ and $5.17 \times 10^{-31}$, respectively.
In Fig.~\ref{fig:basin_zoom_figure} we show basins on every fourth bisection interval, so two consecutive plots represent magnification by a factor of $2^4 = 16$. 
We show only those intervals with length $\ge 2^{-80} \cdot \Delta \approx 8.27 \times 10^{-26}$.  The magnification plots for $P_1$ demonstrate the existence of fine structure down to the smallest scale resolvable at the limit of our quadruple-precision numerics.  Upon magnification of the narrow beige strip on the third-to-last interval (by a factor of $16$), we find even narrower green and red strips around it. The green strip is identified by the bisection process only after five more bisection iterations beyond the last interval shown in Fig.~\ref{fig:basin_zoom_figure}, when the bisection interval is of length $2.58 \times 10^{-27}$. Thus, the fractality length scale for $P_1$ (with $\Delta = 0.1$ and resolution $1.29 \times 10^{-27}$) is $\ell = 2.58 \times 10^{-27}$. In contrast, for the cross section through $P_2$, the plots indicate that the boundary becomes simple at a scale well above the numerical resolution, with the narrowest observed strip of basin found on the bisection interval of length $5.68 \times 10^{-15}$ (the green part in the middle of the 12th plot in Fig.~\ref{fig:basin_zoom_figure}). 
With two more iterations applied to this interval, we have an interval of length $1.42 \times 10^{-15}$ (not shown), and the midpoint of that interval belongs to the green strip.  Since this is the last time this situation occurs, the fractality length scale for $P_2$ is $\ell = 1.42 \times 10^{-15}$ (with $\Delta = 0.1$ and  resolution $5.17 \times 10^{-27}$).

\section{Distribution of fractality measures}
\label{sec:dist_frac_meas}

\begin{figure}[t]
\begin{center}
\includegraphics[width=\columnwidth]{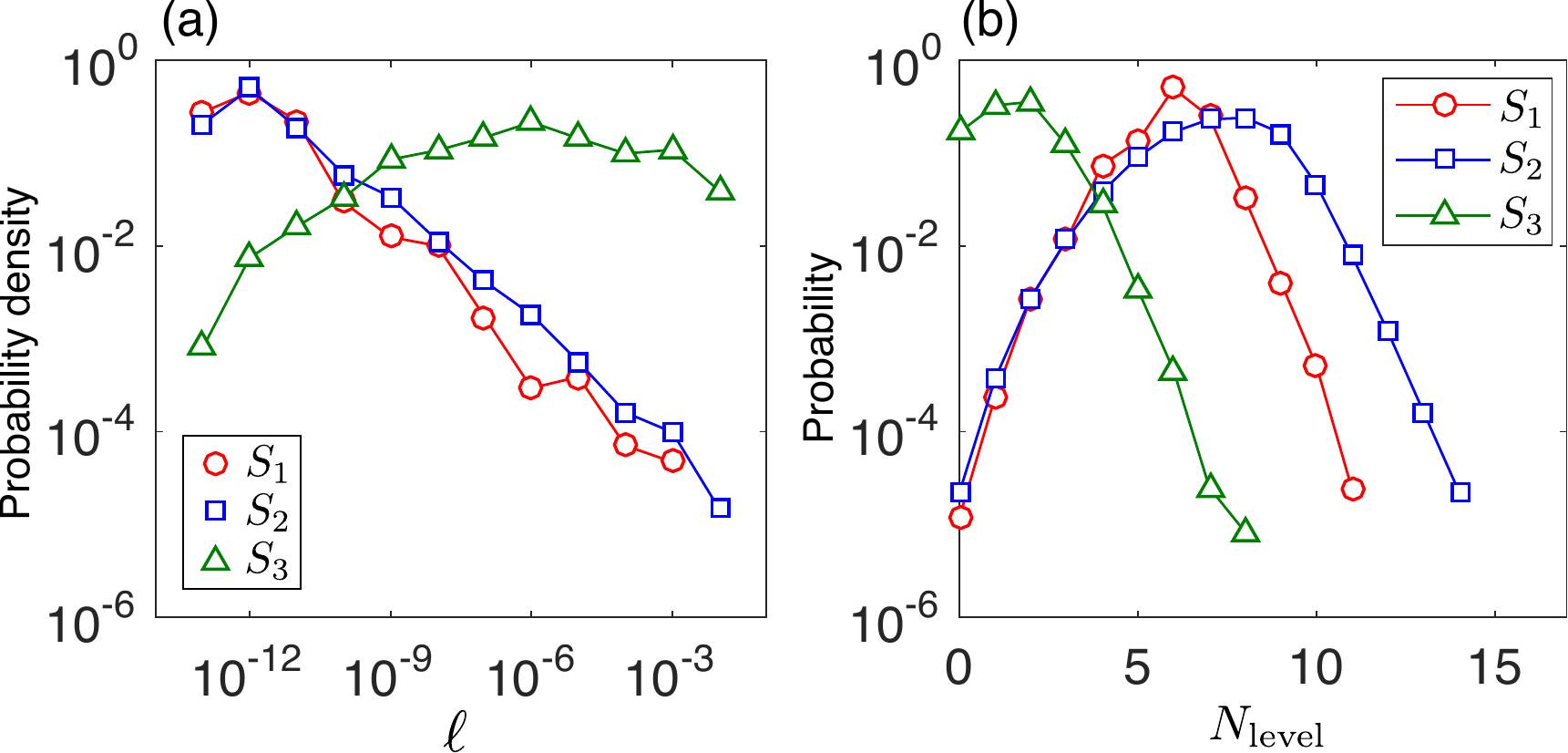}
\vspace{-9mm}
\end{center}
\caption{Distribution of fractality measures for the roulette system computed from the $2^{20} \approx 1.05 \times 10^6$ grid points used in Figs.~\ref{fig:basin-boundaries}(d)--\ref{fig:basin-boundaries}(f).
(a)~Estimated probability density function for the fractality length scale $\ell$. (b) Estimated probability mass function for the construction level $N_\text{level}$.}
\label{fig:res_level_hist_fig}
\end{figure}

Figure~\ref{fig:res_level_hist_fig}(a) shows the probability density functions for the fractality length scale $\ell$ estimated using the same set of line segments used to generate Figs.~\ref{fig:basin-boundaries}(d)--\ref{fig:basin-boundaries}(f).
Note that $\ell$ is finite and broadly distributed above the numerical precision even for the true fractal in the case of $S_1$, since the next smaller scale at which finer basin structure is observed can be below the level of numerical resolution.  For $S_1$, we have $\ell < 10^{-10}$ for most boundary points (more than 95\% of the approximately $10^6$ line segments used), which is consistent with the true fractality of the boundaries.  For both $S_2$ and $S_3$, the boundaries are finite-scale fractals; however, $\ell$ for $S_3$ is larger than $10^{-9}$ in more than 90\% of the boundary points, indicating that the simple boundaries can almost surely be observed after zooming in a few times, while $\ell$ for $S_2$ is $< 10^{-10}$ in about 93\% of the cases, suggesting that the simple boundaries at small scales are mostly hidden behind numerical round-off errors.

As a measure to quantify the extent to which the boundaries exhibit the Wada property, we define the \textit{construction level} $N_\text{level}$ through the same bisection process we used to define $\ell$.  Rather than using interval length, however, $N_\text{level}$ is defined as the number of times the same situation (i.e., the end points and the midpoint all belonging to different basins) occurs in the process.  Note that being able to continue the bisection process indefinitely implies that points belonging to all three basins can be found in an arbitrarily small interval, indicating the Wada property.  Thus, $N_\text{level}$ can be interpreted as the depth of the Cantor-set construction levels observed by the bisection procedure, and hence a quantitative measure of the Wada property.  Figure~\ref{fig:res_level_hist_fig}(b) shows the probability distributions of $N_\text{level}$ estimated  from the set of line segments used for Figs.~\ref{fig:basin-boundaries}(d)--\ref{fig:basin-boundaries}(f).
The construction levels for $S_3$ are relatively small, as expected for finite-scale fractals.  However, $N_\text{level}$ for $S_2$ is significantly larger than for $S_1$ on average, which is the opposite of what one might expect, since the boundaries form a finite-scale fractal for $S_2$ and a true fractal for $S_1$; however, this is consistent with the observation that the complexity of the boundaries for $S_2$ in Fig.~\ref{fig:basin-boundaries} appears to be higher than that for $S_1$.

\begin{figure}[t]
\begin{center}
\includegraphics[width=\columnwidth]{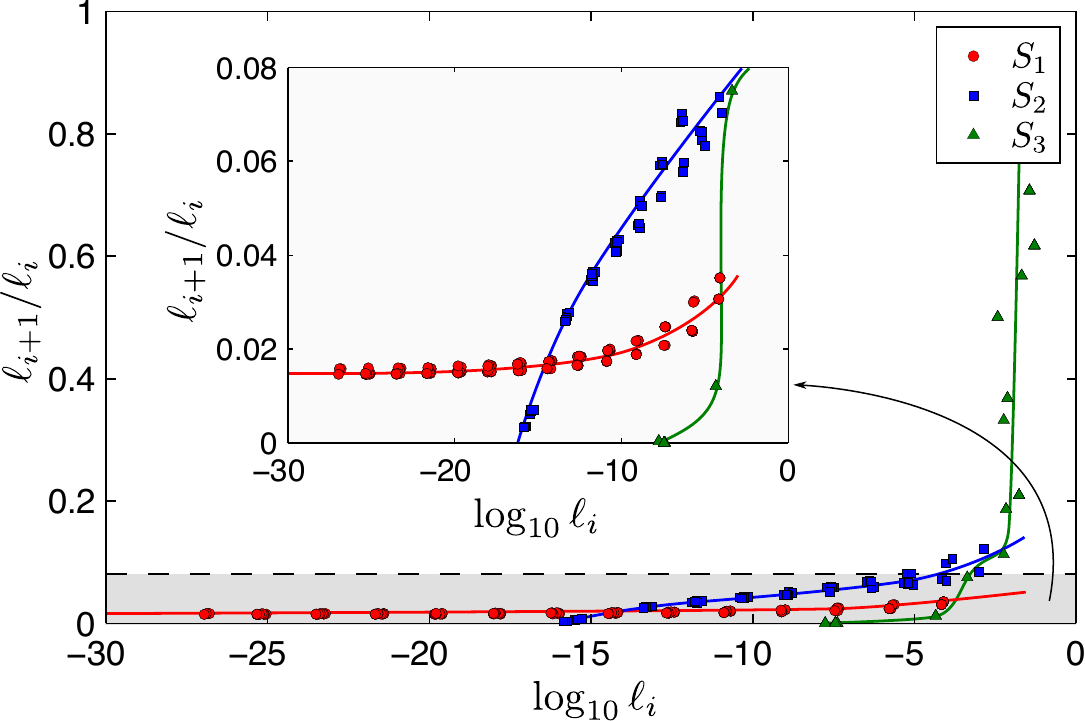}
\vspace{-9mm}
\end{center}
\caption{Ratio $\ell_{i+1}/\ell_i$ of consecutive basin interval lengths for the roulette system as a function of $\ell_i$.
The inset shows a magnification of the region shaded in gray.
For each shape of the roulette surface ($S_1$, $S_2$, or $S_3$), ten realizations of the (random) process described in Appendix~\ref{sec:basin_int_length} are superimposed.
The curves are to guide the eyes.}
\label{fig:interval_length}
\end{figure}

\section{Length of basin intervals}
\label{sec:basin_int_length}

To account for scales below $10^{-14}$ in estimating the fractal dimension, we directly measure the length of the basin intervals in quadruple precision on a (one-dimensional) line segment in the phase space (which is denoted by $I_0$ and has length $\ell_0$) using an iterative procedure.  In the $n$th iteration of this procedure we apply the following steps to the segment (interval) $I_{n-1}$ from the previous iteration:
\begin{enumerate}
\item Divide $I_{n-1}$ into $M=100$ subintervals of equal length, giving $M+1$ uniformly spaced end points, which we denote $x_0, \ldots, x_M$.  Determine the basin to which each $x_i$ belongs by computing the trajectory starting from $x_i$.  
\item Identify $i_1$, the first index $i$ for which $x_{i}$ belongs to a different basin than the one to which $x_{i+1}$ belongs.  Similarly, identify $i_2$, the last index $i$ for which $x_i$ belongs to a different basin than the one to which $x_{i-1}$ belongs.  To ensure that the points $x_0, \ldots, x_M$ capture the basin structure in the interval $I_n$ at a sufficiently high resolution, we check whether $(x_{i_2} - x_{i_1})/(x_M - x_0) \ge 0.9$ holds.  If it does, proceed to step 3.  Otherwise, redefine $I_{n-1} \equiv [x_{i_1},x_{i_2}]$ (to zoom in on the boundary points) and go back to step 1.
\item Among $x_0, \ldots, x_M$, identify the largest set of consecutive points that belong to the same basin, but containing neither $x_0$ nor $x_M$.  The interval $[x_{i_3},x_{i_4}]$, where $x_{i_3}$ and $x_{i_4}$ are the first and last points in the set, respectively, defines the basin interval to be removed in this iteration of a ``Cantor-set construction'' of the basin boundaries.  The removal of this interval leaves two subintervals, one of which is chosen randomly with equal probability as the next interval, denoted $I_n$.
\end{enumerate}
Iterating this procedure starting with $I_0$, we obtain a sequence of intervals, $I_0, I_1, \ldots$, having length $\ell_0, \ell_1, \ldots$, respectively.  

For the standard Cantor set in which two equal-length intervals are left after removing a middle part of the interval in each iteration, the ratio $\ell_{i+1}/\ell_i$ for the sequence generated by the above procedure is constant (i.e., independent of $i$), and the fractal dimension of the set is given by $D = - \frac{\ln 2}{\ln(\ell_{i+1}/\ell_i)}$.  Thus, in general, if we apply the procedure above to a one-dimensional cross section of fractal basin boundaries (e.g., for $S_1$), we expect $\ell_{i+1}/\ell_i$ to approach a positive constant, which is a behavior equivalent to that of the Cantor set whose dimension is given by Eq.~\eqref{eqn:dim-interval-len}.
This is indeed observed numerically for the roulette system with shape $S_1$ [starting from $I_0$ defined by $3.2 \le v_0 \le 3.3$ and $\theta_0 = 0.8$ in the $(v,\theta)$ projection of the phase space], as shown in Fig.~\ref{fig:interval_length}, where we have $\ell_{i+1}/\ell_i \approx 0.015$ for the smallest values of $\ell_i \approx 10^{-27}$, corresponding to $D \approx 0.17$.

For finite-scale fractal boundaries (e.g., for $S_2$ and $S_3$), the basin structure is simple below a certain length scale.  This means that for some sufficiently large $n$, the interval $I_{n-1}$ in step 1 consists of two basin intervals with one simple boundary point in between.  This leads to an infinite loop repeating steps 1 and 2, which represents an endless sequence of magnification by a factor of $1/M$, zooming in on the boundary point.  In this case the procedure described above gives a finite sequence of intervals, $I_0, I_1, \ldots, I_N$, for some $N$.  Figure~\ref{fig:interval_length} demonstrates this behavior for both $S_2$ and $S_3$, showing no intervals with $\ell_i < 1.0 \times 10^{-16}$ for $S_2$ (starting from $I_0$ defined by $2.4 \le v_0 \le 2.5$ and $\theta_0 = 1.8$) and no intervals $\ell_i < 1.7 \times 10^{-8}$ for $S_3$ (starting from $I_0$ defined by $2.7 \le v_0 \le 2.8$ and $\theta_0 = 0.8$).  This is consistent with our theoretical result that $D=0$ for $S_2$ and $S_3$.


\end{document}